\documentclass[fleqn,twoside]{article}
\usepackage{espcrc2}

\usepackage{graphicx}
\usepackage[figuresright]{rotating}

\newcommand{\AmS}{{\protect\the\textfont2
  A\kern-.1667em\lower.5ex\hbox{M}\kern-.125emS}}

\title{Multiplicity fluctuations in high energy hadronic and nuclear
collisions} 

\author{M.Rybczy\'nski\address[a]{Institute of Physics,
        \'Swi\c{e}tokrzyska Academy, Kielce, Poland}
        \thanks{email: mryb@pu.kielce.pl},
        Z.W\l odarczyk\addressmark[a]\thanks{e-mail: wlod@pu.kielce.pl},
        O.V.Utyuzh\address[b]{The Andrzej Soltan Institute for
        Nuclear Studies, Nuclear Theory Department, Warsaw, Poland}
        \thanks{email: utyuzh@fuw.edu.pl}
        and G.Wilk\addressmark[b]\thanks{e-mail: wilk@fuw.edu.pl}}

\begin{document}

\begin{abstract}
The showers of cosmic rays entering the Earth's atmosphere are main
sources of information on cosmic rays and are also believed to
provide information on elementary interactions at energies not 
accessible to accelerators. In this context we would like first to
remind the role of inelasticity $K$ and elementary cross section
$\sigma$ and then argue that similar in importance are fluctuations
of different observables. The later will be illustrated by
multiplicity fluctuations in hadronic and nuclear collisions.
\vspace{1pc}
\end{abstract}

\maketitle

\section{INTRODUCTION}

The EAS (Extensive Air Showers) initiated by cosmic rays (CR)
entering the Earth's atmosphere are our main tool to investigate the
CR as such and they are believed to remain our only possible window
to look at the elementary interactions at energies not accessible to
accelerators \cite{CR}. Agreeing fully with the first part of this
statement we would like to remind here that EAS are themselves
macroscopic stochastic processes of variable length and this fact
sometimes leads to results, which are unexpected from the elementary
particle interactions point of view, the apparent intermittency
seen there and connected solely with stochasticity of EAS only
\cite{INTERM} being the best example. Other example worth to mention
at this point is the apparent self-organized character of EAS
depending in a visible way on their length and clearly seen in data
\cite{SOCEAS}. Both phenomena do not depend on details of elementary
interactions used in describing EAS. Actually, as it was widely
appreciated some time ago \cite{SWWW} and is being rediscovered also
at present \cite{Know}, out of numerous parameters of models of
elementary interactions used in MC codes describing production and
development of EAS the most important are: inelasticity $K$ and
elementary cross section $\sigma$ (actually, the hadron - air nucleus
cross section). Out of these two, the inelasticity $K$, i.e., the
fraction of energy used for production of secondaries and therefore
not available for the subsequent interactions in developing EAS, is
nowadays not really a parameter as it used to be before (cf.
\cite{SWWW}) but rather a number which must be reproduced by
combination of all parameters used in the present models
\cite{Know}\footnote{Actually $K$ appears usually together with
$\sigma$, therefore, as we have shown in \cite{SWWW}, one needs at
least two independent different experiments (measuring quantities in
which $K$ enters in different way) to be able to estimate $K(s)$ from
experiment in a model independent way. For other possibilities see
\cite{KWW} and discussion below.}. 

\section{FLUCTUATIONS}

However, there are some data (cf. \cite{Surprise}) which seem to
demand more information. In particular, in \cite{Fluc} we have shown
that the phenomenon of the so called {\it long-flying-component} can
be most naturally explained by allowing for the fluctuations in the
cross section (which at that time were also widely discussed in the
literature in other context). This conjecture was later quantify by
using the so called nonextensive statistical approach (essentially
corresponding to using Tsallis entropy characterized by parameter $q$
such that for $q \longrightarrow 1$ it becomes the usual
Boltzmann-Gibbs entropy) \cite{neq}. In such approach parameter $q$
measures, in a sense, the amount of fluctuations. For the purpose of
this presentation it is enough to say that whenever in the
exponential formula, $\exp( - X/\Lambda)$, parameter $1/\Lambda$
fluctuates according to gamma distribution then one should use
instead expression\footnote{See \cite{interq} for details.
Generalization of this approach to other type of fluctuations is
discussed in \cite{Beck}.} 
\begin{equation}
\exp_q( - X/\Lambda_0) = [1-(1-q)X/\Lambda]^{1/(1-q)} \label{eq:eq}
\end{equation}
where $1/\Lambda_0 = \left\langle 1/\Lambda \right\rangle$ and 
$q = \left\langle \left(1/\Lambda\right)^2 \right\rangle/
\left\langle 1/\Lambda \right\rangle^2$
(averages are over the above gamma distribution). We first applied
this approach to the above mentioned data on long flying component
\cite{LFC} and found that fluctuations they correspond to are given
by $q=1.3$. 
\vspace{-5mm}
\begin{figure}[h]
\setlength{\unitlength}{1cm}
\begin{picture}(8.0,3.5)
\includegraphics{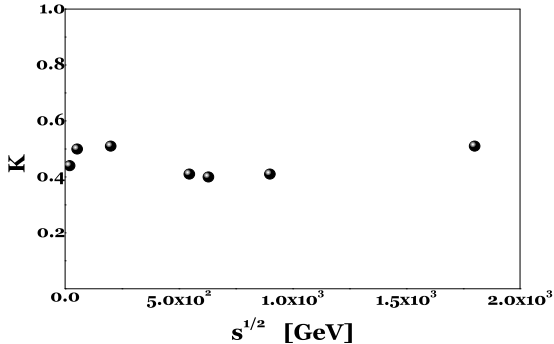} 
\end{picture}
\vspace{-7mm}
\caption{Results on the energy dependence of inelasticity $K$ from
$pp$ and $\bar{p}p$ data (see \cite{Kq} for details).} 
\end{figure}
\vspace{-17mm}
\begin{figure}[h]
\setlength{\unitlength}{1cm}
\begin{picture}(8.0,3.5)
\includegraphics{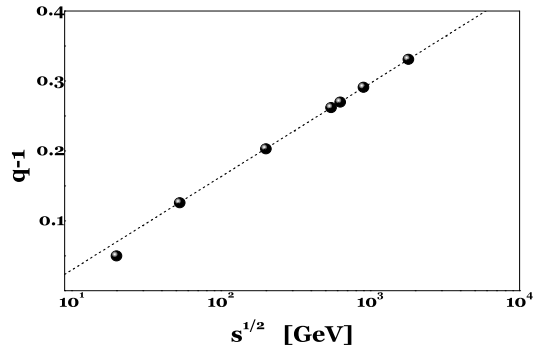} 
\end{picture}
\vspace{-5mm}
\caption{Results on the energy dependence of fluctuations as given by
$q=q(s)$ (points) compared with the best fit to energy dependence of
parameter $1/k$ of Negative Binomial multiplicity distribution (line,
see text and \cite{Kq} for details).}
\end{figure}
\vspace{-5mm}

In Figs. 1-3 we provide results on inelasticity $K$ and fluctuations
seen in the rapidity distributions, which originate from the
multiplicity fluctuations \cite{Kq,Others}. They were obtained using
nonextensive version of the information theory approach to
multiparticle production processes based on the Tsallis $q$-entropy.
Given the mean multiplicity $\langle n\rangle$ and mean transverse
momentum $\langle p_T \rangle$ one gets the most probably and least
biased rapidity distribution, being of the $\exp_q(...)$ type, i.e.,
depending on the nonextensivity parameter $q$, which further depends
only on the amount of energy available for production of secondaries,
$W=K\cdot \sqrt{s}$, i.e., on the inelasticity $K$. In this way
inelasticity $K$ and nonextensivity parameter $q$ specifying entropy
used as measure of information are the only parameters to be fitted.
Fig. 1 shows the energy dependence of inelasticity, $K=K(s)$, whereas
Fig. 2 the energy dependence of fluctuations as given by $q=q(s)$.
In the case when one knows rapidity distributions also for a given
multiplicity (or, at least, for some multiplicity bins) this method
allows to deduce also inelasticity distribution, as can be seen in
Fig. 3 where it was done for $\sqrt{s} = 200$ GeV (black symbols,
$K=0.52$) and $900$ GeV (open symbols, $K=0.38$). They can be fitted
either by gaussians (full lines) or, better, by lorentzian (dotted
lines) shapes.
\begin{figure}[h]
\setlength{\unitlength}{1cm}
\begin{picture}(8.0,4.2)
\includegraphics{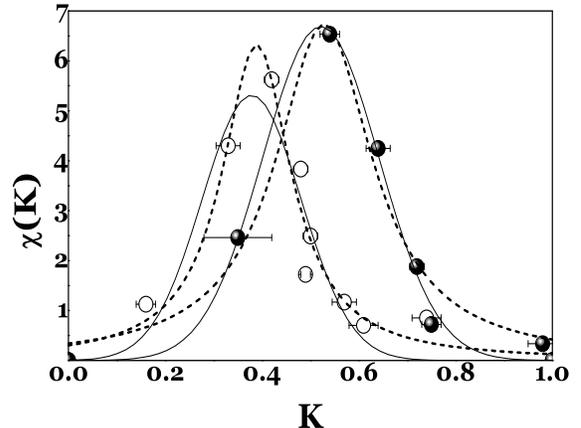} 
\end{picture}
\vspace{-7mm}
\caption{Examples of inelasticity distributions $\chi(K)$ obtained
using the available UA5 data on rapidity distributions in
multiplicity bins (see \cite{Kq} for details).}
\end{figure}
\vspace{-5mm}

In what follows we shall, however, concentrate on fluctuations
described by the parameter $q$ shown in Fig. 2. We argue that these
fluctuations are due to fluctuations of mean multiplicity used as our
input when deducing inelasticity $K$. Namely, when there are only
statistical fluctuations in the hadronizing system one expects
Poissonian form of the corresponding multiplicity distributions. The 
existence of intrinsic (dynamical) fluctuations means then that one
allows mean multiplicity $\bar{n}$ to fluctuate. In the case when
these fluctuations are given by gamma distribution with normalized
variance $D(\bar{n})$ then, as a result, one obtains the Negative
Binomial multiplicity distribution, which depends on two parameters:
the mean multiplicity $\langle n\rangle$ and the parameter $k$ ($k\ge
1$) affecting its width,
\begin{equation}
\frac{1}{k}\, =\, D(\bar{n})\, =\,
\frac{\sigma^2\left(\bar{n}\right)}{\langle \bar{n}\rangle^2} .
\label{eq:D} 
\end{equation}
That is because (see \cite{Shih}):
\begin{eqnarray}
P(n) &=& \int_0^{\infty} d\bar{n}
             \frac{e^{-\bar{n}}\bar{n}^n}{n!}\cdot  
               \frac{\gamma^k \bar{n}^{k-1} e^{-\gamma
              \bar{n}}}{\Gamma (k)} = \nonumber\\
     &=&  \frac{\Gamma(k+n)}{\Gamma (1+n) \Gamma (k)}\cdot 
          \frac{\gamma^k}{(\gamma +1)^{k+n}} \label{eq:PNBD}
\end{eqnarray}   
where $\gamma = \frac{k}{\langle \bar{n}\rangle}$. According to our
philosophy it is therefore natural to describe these fluctuations by
the nonextensivity parameter $q$ assuming that $D(\bar{n}) = q-1$,
i.e., that $q = 1 + 1/k$. However, in the nonextensive approach of
this type $q$ is limited by $q < 3/2$ \cite{Others}. This condition
would then impose constraint on the amount of fluctuation as it
corresponds to $k<2$, the saturation limit, which from the naive
extrapolation of tendency presented in Fig. 2 would appear at
energies $\sim 33$ TeV (or $E_{LAB} = 0.5\cdot 10^{18}$ eV) range. 
Notice that this energy range of the ultra high energetic cosmic rays in
which effects connected with the GZK cut-off starts to be important
\cite{CR}. It means therefore that the knowledge of fluctuations will
most probably turn out to be as important as that of inelasticity and
cross section.

We would like to add to this one more observation. In Fig. 4 we have
plotted experimental data show on mean charged multiplicity $\langle
n_{ch}\rangle$ at different energies versus corresponding total
inelastic cross section $\sigma$ (Fig. 4a) and the estimated volume
of the interaction region $V$ (taken as being proportional to
$\sigma^{3/2}$). The linearity of $\langle n_{ch}\rangle$ and
$\sigma$ is remarkable. This suggest that fluctuations of the multiplicity
distributions should also be connected with fluctuations of the cross
section but at the moment there are no investigations in this
direction\footnote{Although constraints on the production cross
section imposed by the observed behaviour of multiplicity
distributions have been investigated already long time ago 
\cite{SIGPN}, this subject is not pursued at moment. In models so far
there is problem with $\langle n_{ch}\rangle$ at energies of interest
to CR as predictions vary in unacceptable way \cite{Showersim}.}.
\vspace{-8mm}
\begin{figure}[h]
\setlength{\unitlength}{1cm}
\begin{picture}(8.0,8.3)
\includegraphics{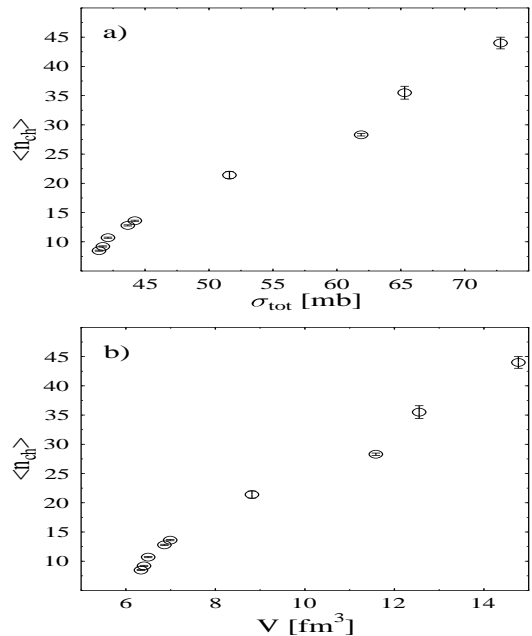} 
\end{picture}
\vspace{-10mm}
\caption{$(a)$ Experimentally observed correlation between $\langle
n_{ch}\rangle$ and $\sigma$. $(b)$ The same but with $\sigma$
replaced by the volume of interaction estimated as being $\propto
\sigma^{3/2}$.} 
\end{figure}
\vspace{-10mm}

\section{SUMMARY}

To summarize: phenomenon of EAS is macroscopic and so complicated
that it depends only on a small number of characteristics of
elementary interactions. It seems that they are: inelasticity $K$,
cross section $\sigma$ and its fluctuations. We have argued that
although we do not measure directly the later we measure fluctuations
of multiplicity distributions, which influence such observables as
rapidity distributions. It is argued that they should also provide us
with some estimations of the fluctuations of the cross section,
albeit we do not yet know the respective formulas.
\vspace{-2mm}
\begin{figure}[h]
\setlength{\unitlength}{1cm}
\begin{picture}(8.0,5)
\includegraphics{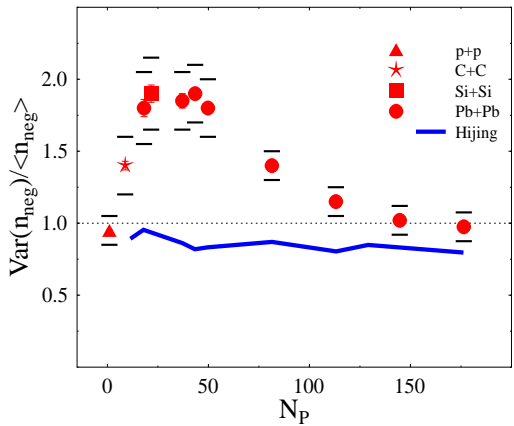} 
\end{picture}
\vspace{-10mm}
\caption{Results on the nonmonoticity of fluctuations represented by
$Var(n_{neg})/\langle n_{neg}\rangle$ as function of number of 
participating nucleons (cf. \cite{Nonmon}).}
\end{figure}
\vspace{-6mm}
However, there are already some data, which warn us that to get such
connection could be very difficult task. Namely, in Fig. 5 we see
clearly that such fluctuations measured in nuclear collisions are
nonmonotonic function of the number of participating nucleons, which
in turn is proportional to the volume of interaction
\cite{Nonmon}\footnote{As shown in \cite{RWW} for nuclear collisions 
fluctuations in $n$ are directly connected with the equation of state
parameter $c=p/\varepsilon$ and $q$: $Var(n) = (q-1)\cdot
\langle n\rangle^2/(1+c)^2 \simeq 0.28\cdot \langle n\rangle^2$ for
$c=1/3$ and limiting value of $q=1.5$.}. This could suggest that
monotonicity observed in behaviour of first moments (cf. Fig. 4) could
be lost when going to higher moments.


\begin{thebibliography}{9}

\bibitem{CR} See, for example, talks by: K. Shinozaki (AGASA), S.
             Westerhoff (HIRES), K.H. Kampert (AUGER) and A. Haungs
             (KASKADE), these proceedings. 

\bibitem{INTERM} G. Wilk and Z. W\l odarczyk, J. Phys. G19 (1993) 761.

\bibitem{SOCEAS} M. Rybczy\'nski, Z. W\l odarczyk and G. Wilk,
                 Nucl. Phys. B Proc. Suppl. {\bf 97} (2001) 81.

\bibitem{SWWW} Yu.M. Shabelski, R.M. Weiner, G. Wilk and 
               Z. W\l odarczyk, J. Phys. G18 (1992) 1281.

\bibitem{Know} Cf., for example, contributions by H.J. Drescher, S.
               Ostapchenko and A. Haungs, these proceedings.

\bibitem{KWW} G. Wilk and Z. W\l odarczyk, Phys. Rev. D59 (1999) 
              014025 and Nucl. Phys. B Proc. Suppl. 75A (1999) 171.

\bibitem{Surprise} G. Wilk and Z. W\l odarczyk, Acta Phys. Polon. B27
                   (1996) 2649.

\bibitem{Fluc} G. Wilk and Z. W\l odarczyk, Phys. Rev. D50 (1994) 
               2318.

\bibitem{neq} Cf., for example, G. Wilk and Z. W\l odarczyk, 
              Physica A305 (2002) 227.

\bibitem{interq} G. Wilk and Z. W\l odarczyk, Phys. Rev. Lett. 84 
                 (2000) 2770 and Chaos, Solitons and Fractals 13 
                 (2002) 581.

\bibitem{Beck} C. Beck and E.G.D. Cohen, Physica A322 (2003) 267.

\bibitem{LFC} G. Wilk and Z. W\l odarczyk, Nucl. Phys. B Proc. Suppl.
              75A (1999) 191.

\bibitem{Kq} F.S. Navarra, O.V. Utyuzh, G. Wilk and Z. W\l odarczyk,
             Phys. Rev. D67 (2003) 114002.

\bibitem{Others} F.S. Navarra, O.V. Utyuzh, G. Wilk and Z. W\l odarczyk,
                 Nuovo Cim. 24C (2001) 725, Physica A340 (2004) 467;
                 hep-ph/0312136 (to be published in Physica A (2004))
                 and hep-ph/0312166 (to be published in Nukleonika
                 (2004)). Cf. also: M. Rybczy\'nski, Z. W\l odarczyk
                 and G.Wilk, Nucl. Phys. B Proc. Suppl. 122 (2003)
                 325.

\bibitem{Shih} P. Carruthers and C.S. Shih, J. Mod. Phys. A2 (1986) 
               1447.

\bibitem{NB} C. Geich-Gimbel, Int. J. Mod. Phys. A4 (1089) 1527.
                                  
\bibitem{SIGPN} C. Novero and E. Predazzi, Nuovo Cim. A63 (1981) 129;
                H. Yokomi, Prog. Theor. Phys. 55 (1976) 2023 and 
                Phys. Rev. Lett. 36 (1976) 924.

\bibitem{Showersim} J. Knapp et al., Astropart. Phys. 19 (2003) 77.

\bibitem{Nonmon} M. Rybczy\'nski (for NA49 Collab.), nucl-ex/0409009;
                 St. Mr\'owczy\'nski, M. Rybczy\'nski and Z. 
                 W\l odarczyk, nucl-th/0407012 (to appear in Phys.
                 Rev. C (2004)).

\bibitem{RWW} M. Rybczy\'nski, Z. W\l odarczyk and G. Wilk, Acta 
              Phys. Polon. B35 (2004) 819.

\end{thebibliography}
\end{document}